# Feature Learning for Stock Price Prediction Shows a Significant Role of Analyst Rating

**Jaideep Singh** and **Matloob Khushi** *

School of Computer Science, Faculty of Engineering, The University of Sydney, NSW 2006, Sydney Australia
* Correspondence: mkhushi@uni.sydney.edu.au

**Abstract:** Efficient Market Hypothesis states that stock prices are a reflection of all the information present in the world and generating excess returns isn't possible by merely analysing trade data which is already available to all public. Yet to further the research rejecting this idea, a rigorous literature review was conducted and a set of 5 technical indicators and 23 fundamental indicators was identified to establish the possibility of generating excess returns on the stock market. Leveraging these data points and various classification machine learning models, trading data of the 505 equities on the US S&P500 over the past 20 years was analysed to develop a classifier effective for our cause. From any given day, we were able to predict the direction of change in price by 1% up to 10 days in the future. The predictions had an overall accuracy of 83.62% with a precision of 85% for buy signals and a recall of 100% for sell signals. Moreover, we grouped equities by their sector and repeated the experiment to see if grouping similar assets together positively effected the results but concluded that it showed no significant improvements in the performance – rejecting the idea of sector-based analysis. Also, using feature ranking we could identify an even smaller set of 6 indicators while maintaining similar accuracies as that from the original 28 features and also uncovered the importance of buy, hold and sell analyst ratings as they came out to be the top contributors in the model. Finally, to evaluate the effectiveness of the classifier in real-life situations, it was backtested on FAANG equities using a modest trading strategy where it generated high returns of above 60% over the term of the testing dataset. In conclusion, our proposed methodology with the combination of purposefully picked features shows an improvement over the previous studies, and our model predicts the direction of 1% price changes on the 10$^{th}$ day with high confidence and with enough buffer to even build a robotic trading system. The code is available from http://mkhushi.github.io/

## 1 Introduction

There has been extensive research to predict a firm's performance [1,2], with some of the earliest studies beginning in the 1980s. Yet, as expressed by Kiang [3], the progress in this field has been limited primarily because of lack of willingness on behalf of security analysts to share their knowledge and their inability to articulate their knowledge in unambiguous and simple terms for the domain to be taken forward. Moreover, a popular idea, the Efficient Market Hypothesis by Malkiel and Fama [4] states that prices of stocks are informationally efficient i.e. the stock prices reflect all information available out there and it is not possible to predict stock prices based on the trading data and nor is it possible to obtain excess returns by exploiting any predictability of prices. Yet, on the other hand, Cheung et al [5] and Pesaran et al [6] argued and established that to execute profitable trading, forecasting just the direction is enough.

Usually when trading, to evaluate the price of a security one of the 2 strategies is adopted: the fundamental analysis [7] or the technical analysis [8]. Fundamental analysis determines a security's value by evaluating the underlying factors that affect a company's current business and its future prospects whereas technical analysis revolves around analysing statistical trends gathered from trading activity, such as price and volume movement and to identify windows of opportunities to invest. Both of these techniques approach the task from quite different angles and often are chosen one over the other depending upon the context of many factors including the market, equity in consideration, investment period such as long term or short term etc [9,10]. Typically, studies were carried to check the feasibility of one of those techniques [7,11,12] but in the study by R. A. Kamble [13], it was argued that technical indicators are not sufficient for long-term prediction of stock while fundamental and technical data together can make accurate long-term stock prediction possible. Thus, implying that no strategy serves as a silver bullet, and most strategies are developed and optimized specifically to work only in certain kind of markets or with certain kinds of equities. Thus, leaving a wide gap open and leaving room for further research in this domain.

Parallelly, with improvements in machine learning and the availability of increased computational resources, many machine learning algorithms like Neural Networks [14-18], Random Forests [13], Bayesian Networks [19], hybrid models [20] and even very specific, optimised ML models [21] have all been applied in various scenarios to better predict the nature of the markets with a fair degree of success. Often, Support Vector Machines (SVM) have been leveraged for technical analysis to perform regression operations [16]; for fundamental analysis, Neural Networks and its variations have been used to capture patterns and relationships by classifying the stocks based on the selected accounting indicators and macroeconomic factors. Despite these ML models being successful in predicting stock price trends [15] up to a certain level, as evident, they have been leveraged in very specific scenarios for very specific functions – making them fit primarily for the role they were chosen for.

Thus, the key question in algorithmic trading is to identify how to define a set of rules on historical stock data like price, volume, market cap etc. as well as other publicly available information like a company's annual reports, expert recommendations, overall market condition to accurately predict market behaviour and correctly identify trading opportunities to generate excess returns. In this study, we collect features that were identified in various studies, study them, apply them and evaluate the possibility of predicting stock price movement whilst leveraging the advances in machine learning methods.

## 2 Related Work

### 2.1 Predicting an equity's performance

Inceptive research in fundamental analysis had mainly focused on using statistical methods to relate the performance of a stock with measurable parameters like financial ratios, and then make suggestions to buy or sell, as the values of these parameters changed [11,22]. For instance, Ou [7] surveyed 61 accounting descriptors to shortlist 8 descriptors which were then run through various logistic regression models to establish that stock prices behave as if investors revise their expectations of future earnings based on non-earnings annual report numbers of the current year. Whereas Reinganum [22] by selecting merely 8 fundamental indicators was able to pick stock strategically to give returns 2-3 times higher than the S&P500 depending upon the selection strategy.

Various studies [9-11,14,18,23] have established that fundamental analysis by incorporating trading characteristics of the firm, such as the macroeconomic variable, present a possibility of good returns. Particularly, Beneish et al [9] showed that the stocks chosen using such a strategy outperformed the losers by 8.7% to 17.8% over 12 months and Tirea and Negru [23] confirmed this claim by increasing the number of indicators and incorporating inputs on volatility, company management along with fundamental and macroeconomic indicators. Moreover, Anderson [10] established that fundamental signals can have different implications across the life-cycle of a firm i.e. an indicator that is highly correlated to growth in a start-up might loosely or even negatively correlate to success in a medium to a large-sized.

In the context of our study and machine learning, models learn and base their actions on the past, if a situation arises from beyond its domain of learning, by the nature of its design, the model is susceptible to performing really poorly or perhaps even break completely. The model is 'only as good as what it is prepared for'. Yet if the stock predictions are based on fundamental analysis which promotes looking at a security's intrinsic value than the market value, or the charts that represent the market value, the model could be made resilient to an extent. Thus, the amalgamation of the fundamental and technical factors is taken up as a potential solution.

*2.2 Machine learning strategies for good performing stock*

As touched upon earlier, with the advent of Machine learning and increased computational resources this field took a whole different course and since then many different kinds of ML models have been tried and tested in this domain. Though, more importantly, Kryzanowski et. al. [18] reported that a Boltzmann machine (a kind of Artificial Neural network) successfully classified stock returns into Positive-Negative up to 72% based on the selected fundamental indicators and if the classification was changed to Positive-Neutral-Negative then the results weren't as phenomenal but still better than a random guess. Overall, the ability of Neural Networks to perform better than random guess was confirmed when Yu et al. [24] utilized ANNs to select stocks or [17] used it to predict stock returns for Dow Price Index or Trigueiros [8] leveraged a Multilayer Perceptron to test their custom fundamental analysis strategy. In another interesting study by Lam [25], she investigated the ability of neural networks to integrate fundamental and technical analysis to predict the performance of a stock and developed a model averaging a return of 0.25398 which outperformed the minimum benchmark but did not meet the maximum benchmark set at 0.278.

Referring to the application of ANNs across financial markets[26-30] and for both fundamental and technical analysis, Yu et. al. [21] acknowledged the wide application of ANNs but then proceeded to provide evidence that ANNs often suffer from over-fitting and local optimum problems. They also proposed that SVM's based on the principle of structural risk minimization could avoid such problems and employed them to model stock markets. Moreover, they developed a novel Sigmoid function to combine discrete and continuous inputs and later used differential evolution algorithms to optimise the stock selection model to find out that this strategy was significantly more powerful and efficient than other designed benchmarks. Similarly, many other custom and very-tailored approaches were adopted included Huang et al [20] proposing a Fuzzy-GA hybrid which beat the benchmarks and Kiang et. al. [3] developed a custom "distributed knowledge acquisition system" which had dedicated components to perform specific analysis functions and when the output of it all was combined to select stocks, it outperformed NYSE and S&P500 on the parameters of monthly return, the Treynor Index and the Jensen Index. While Faustryjak et. al. [31] used LSTM with statistical analysis of published messages to predict the direction of stock price and Paluch [32] combined technical and fractal indicators for the next day close price of the Polish stock market.

Furthermore, insightful studies with the other kinds of ML models include Kamble [13] utilizing the Random Forest algorithm on selected indicators based on their short term and long-term impact and found that the accuracy of the model was 66.8% for buy signal for short term and 75.8% for long term. Whereas the experiments using Bayesian Networks by Tan et al [19] aiming to model financial ratios of Malaysian plantation stocks around the period of the Great Recession (2007-09) received limited success as the algorithm was completely unable to predict the crash but its accuracy to classify stocks into 'buy' and 'don't buy' improved post the period of decline to 52.94% and 60.71% in 2009 and 2010 respectively. For the authors, this was a success to an extent as despite such a massive market disruption the model was still able to predict stock returns up to a certain accuracy.

Lastly, Dropsy [14] attempted to predict international equity risk premium using linear regression models and neural networks and benchmarked the results of each against the statistical Random walk model. It was argued that linear (Linear Regression) and non-linear forecasts (Neural Network) are superior to random walk forecasts, but interestingly nonlinear forecasts did not significantly outperform linear forecasts and the unexpected result was attributed to the possibility of the neural network suffering misspecification. And G. Iuhasz, M. Tirea, and V. Negru [33] comparing 4 neural networks namely Standard Feed-Forward neural network, Elman and Jordan recurrent neural networks and a neural network evolved with neuro-evolution of augmenting topologies (NEAT) found out that NEAT computes a more accurate result and much faster than the others.

Apart from the variety of the ML models used, different studies have relied on different features and indicators to predict stock prices. For example, in the Australian market, C. Hargreaves and Y. Hao [15] established that by using only a selected few parameters (less than 10) with either one of, a neural network or a decision tree to pick Australian stocks they could get a better yield than Australian Ordinary Index numerically 11.8 times better.

*2.3 Impact of analyst recommendations*

In a nutshell, analyst recommendations [34] are typically buy-hold-sell ratings predicting an equity's future performance and are issued in the public by industry experts. Womack [35] studied the effect of analyst ratings on investor sentiment and observed that stock prices adjusted up to five per cent for changes to buy recommendations or up to 11 per cent for changes to sell recommendations. Taking it a step forward, Barber et al [36] in 2001 investigated if the investors could profit from these publicly available recommendations and they reported that just by adopting the strategy of purchasing 'strong buy' stocks and shorting 'strong sell' stocks they were able to generate excess returns of up to 4% but stated due to a substantial transaction costs the abnormal net returns for these strategies was not reliably greater than zero. Yet, about more than a decade later in 2018, Park et al [37] used the same strategy to produce abnormal returns of 4.7–5.8% per year post accounting for transactions costs during the period of 2001–2016. They attributed this result to decreasing transaction costs and also to their more precise consideration of these costs rather than using a fixed number. Another recent study [38], suggested that it was possible to build profitable strategies based on analysts' recommendations as they identified that the top-rated stocks generally deliver higher returns than the bottom rated stocks.

While these ratings seem to possess the ability to sway the markets, they are known [39-42] to be biased because of these analysts' affiliations with the investment banking realm that generates pressure on them to produce relatively optimistic recommendations regarding affiliated stocks.

Moreover, Chiang and Lin [43] established that there exists a level of herding in the recommendations issued by analysts and this phenomenon is more profound for hard-to-value firms, large firms or firms with high institutional ownership.

Finally, analyst ratings haven't always proven to be correct [44], as in the study by Nam et al [45], they discovered that stocks with average analyst recommendation level of hold or worse on the S&P500 outperformed other stocks over the period 2009 to 2016. However, according to Nam, this result was driven by two factors: the impact of the period of 2009 and 2010, and low-priced stocks. Yet when the same strategy was applied since the beginning of 2011 or when the low-priced stocks were removed, no significant outperformance of these stocks was observed. This presented an interesting result as these supposedly poor performing stocks as described the analysts didn't truly perform as poorly as predicted but rather managed to stay afloat and provided their investors just as many returns as the stocks with a positive rating if not more.

*2.4 Gap in research*

Studies like [7,22] have highly advocated for fundamental analysis and to check for a security's intrinsic value but as witnessed earlier, technical analysis has also shown promising results. Secondly, most studies seem to have overlooked the impact of analyst ratings in the interest of scope. Thirdly, with rapidly increasing capabilities of Machine learning, new and sophisticated methods [3,21] have been used to analyse the markets, each of which has achieved some success in their own way. Lastly, without taking away from any due credit, it seems because of these new tools provided by machine learning to map out trends, the focus has shifted from picking the right indicators for evaluation to merely checking how these tools work in the given scenario [13-15]. Thus, with all these movings parts and studies focusing on very specific use-cases, there is a wide gap to be explored and to be worked in.

From an application perspective, we realised that most articles predicted just the direction of stock movement or recommended in forms that present a high barrier to be implemented in a real trading system. Therefore, we aimed to propose a system which could be used as a real trading strategy and plan to achieve this by bringing the above-mentioned factors together and focusing on selecting the best possible indicators from published studies covering both the fundamental and technical realms that could have the maximum impact on a security's price movement and by leveraging the latest advances in machine learning [46]. To meet the requirements of this objective we also carried out backtesting of our classifier and evaluated the profits made along the way as explained in the sections below. Figure 1 explains our analysis workflow.

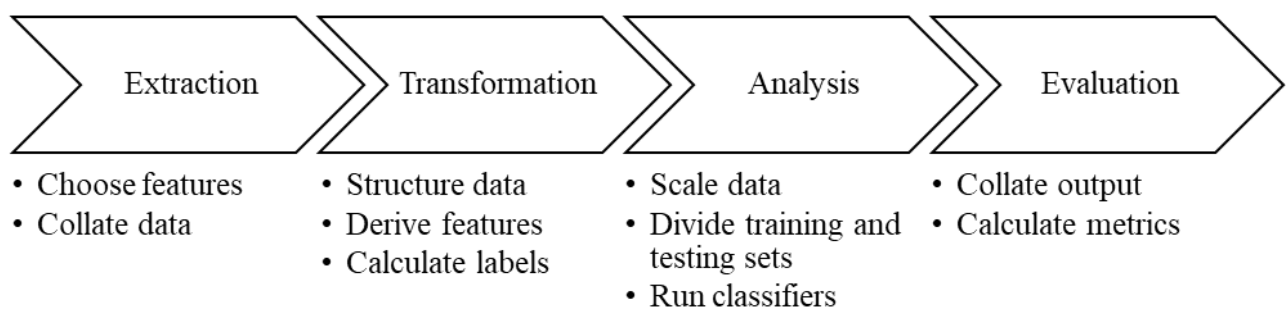

**Figure 1.** Analysis work-flow

## 3 Experiment Design

The entire experimentation process can be divided into 4 stages as depicted in Figure 1. The data focused on the 505 stocks that formed the S&P 500 index and covered the past 20 years until the 4th of April 2019. For these equities, a set of 28 features was identified, collected and structured after a thorough analysis of previous literature as summarised in Table 1 and they ranged from fundamental indicators to technical indicators to analyst recommendations.

**Table 1.** Features and indicators used in this study. * BEst ratings (BEst EPS, BEst CAPEX etc) refer to ratings and numbers calculated and issued by Bloomberg's internal division, Bloomberg Estimate.

| Type | Indicator | Symbol on Bloomberg | Studies |
|---|---|---|---|
| Fundamental Indicators | PE Ratio | PE_RATIO | [5], [9], [11], [17], [21], [24] |
|  | PB Ratio | PX_TO_BOOK_RATIO | [1], [19]–[21] |
|  | EPS * | BEST_EPS | [10], [17], [18], [23], [24] |
|  | EPS Low * | BEST_EPS_LO | " |
|  | EPS High * | BEST_EPS_HI | " |
|  | CAPEX * | BEST_CAPEX | [5], [7], [23] |
|  | CAPEX Low * | BEST_CAPEX_LO | " |
|  | CAPEX High * | BEST_CAPEX_HI | " |
|  | Return on Asset | RETURN_ON_ASSET | [12], [16], [17], [19], [21] |
| Technical Indicators | Volume | PX_VOLUME | [9] |
|  | Market Cap. | CUR_MKT_CAP |  |
|  | Historical Market Cap. | HISTORICAL_MARKET_CAP |  |
|  | Short Interest Ratio | SHORT_INT_RATIO |  |
|  | Short Interest | SHORT_INT |  |
| Analyst Recommendations | # of Analyst Ratings | TOT_ANALYST_REC | [12], [27], [28], [30] |
|  | # of Buy Recs | TOT_BUY_REC | " |
|  | # of Sell Recs | TOT_SELL_REC | " |
|  | # of Hold Recs | TOT_HOLD_REC | " |
|  | Equity Recommendation Consensus | EQY_REC_CONS | " |
|  | Analyst Rating * | BEST_ANALYST_RATING | " |
| Others | Closing Price | PX_OFFICIAL_CLOSE | [13] |
|  | Long-Term Growth * | BEST_EST_LONG_TERM_GROWTH |  |
|  | Target Price for day * | BEST_TARGET_PRICE |  |

| Derived Indicators | %age of Hold Recs | hold_percent |
| --- | --- | --- |
| | %age of Buy Recs | buy_percent |
| | %age of Sell Recs | sell_percent |
| | Standard Deviation in price over 5 days | std_5day |
| | Standard Deviation in price over 10 days | std_10day |

*3.1 Approach for Transformation and Analysis*

Firstly, to achieve the primary aim for this study i.e. to evaluate if a set of technical or fundamental features can be used to predict the price movement of an equity or not, we structured this problem as a classification machine learning problem, similar to the approach adopted by [13,15,19]. Here, we labelled the equity a 'buy' if the price went up in the following days by 1% and 'sell' if the price went down by 1% or if it remained between those 2 thresholds then as a 'hold'. Post this, classification machine learning models could then be run on these to check for an association between the features of an equity and the labels based on price movement for every day.

Secondly, to investigate the price movement for up to 10 days from any given day, each instance in the data (aka row), representing one day, was given 10 labels [day1, day2, day3, day4, day5, day6, day7, day8, day9, day10]. Similar to above, each one of these labels stated 'buy', 'sell' or 'hold' based on the price change on 'day n' in the future compared to the current day. For example, if the label 'day3' of an arbitrary equity on 26/11/11 had a 'buy' value, that would mean that 3 days in the future i.e. 29/11/11 the closing price of the equity was greater than its closing price on 26/11/11 by 1% (Table 2).

Thirdly, to prevent the machine learning models from reporting higher accuracies by incorporating biases and over-reliance on the trends it may pick up, like the Great Economic Recession of 2007-2008 or the Dot-com bubble of 1994 to 2000 or the following Dot-com crash of 2000-2001 where the market showed either meteoric rise or decline for an extended period across the world especially the US, the data should be randomised before forming training and testing sets for the machine learning to gauge nothing but purely the effect of the chosen indicators.

**Table 2**. Criteria for Labels

| Label | Numerical Equivalent | Criterion |
| --- | --- | --- |
| Buy | 2 | $T_n \geq 1.01 * T_0$ |
| Hold | 1 | $0.99 * T_0 < T_n < 1.01 * T_0$ |
| Sell | 0 | $T_n \leq 0.99 * T_0$ |

To execute this approach, post-restructuring, calculating derived features and removing nulls, this transformed dataset is split into 70% and 30% which formed the training set and the testing set respectively and then multiple models were tested, as listed in Table 3. A second iteration of experiments was also run by grouping equities by their sector and classifying each group separately to evaluate the effect of running classifiers on similar kind of equities together.

**Table 3.** List of models used for experimentation

| Model Used | Parameters |
| --- | --- |
| Support Vector Machines | Linear Kernel |
| Naïve Baes | Gaussian |
| | Multinomial |
| Random Forests | No. of trees = 10 |
| Decision Trees | Gini Index |
| | Entropy Information gain |
| K Nearest Neighbours | 5 Neighbours |
| | 7 Neighbours |
| | 9 Neighbours |
| Neural Networks | 2 hidden layers containing 5 and 2 hidden units respectively |

*3.2 Approach for Evaluation*

To do a comprehensive evaluation of the performance of the various models and optimise the utility of the model in a real-world scenario, different metrics were chosen for the individual signals and the overall performance of the model as depicted by Table 4.

For the buy signals, it is important to have a higher precision against recall as we want to make sure if we look at a buy signal produced by our program is truly a buy signal even if we miss out on a few opportunities (i.e. recall, capturing all the signals) to invest here and there.

For the sell signals, the opposite might be true as we would want to know about all the times when we should be selling (i.e. recall, capturing all the signals) because if we don't, then we might be heading towards an impending doom without knowing and can lose all profits there. Thus, it doesn't hurt to sell a bit early (i.e. precision, if captured should be correct) rather than waiting longer and hoping to generate more profits.

As for the hold signal, a low precision would mean one becomes vulnerable in situations when they were supposed to either buy or sell but the classifier tells them to hold making them rigid in terms of their portfolio position – consequently leading to missed investment opportunities or even extreme losses. Similarly, for the recall value, one would want to know about all the hold values because if they don't and are recommended to either buy or sell, then in the best case they become extremely sensitive to the slightest of market changes or worse off compromise their entire portfolio due to an incorrect buy or sell. So, for hold signals, it makes sense to pay attention to both these values and the F-1 score would a great metric for such a situation, accounting for both precision and recall.

**Table 4.** Metrics for different signals

| Entity | Metric |
| --- | --- |
| Buy Signals | Precision |
| Sell Signals | Recall |
| Hold Signals | F1 Score (Micro-Weighted) |
| Overall Model | F1 Score (Micro-Weighted) |

### 3.2.1 Backtesting the classifier

Being a system that needs to be applied in real life, merely calculating theoretical accuracy of its various signals wouldn't mean anything if it is unable to successfully navigate the real world and generate profit. Thus the test dataset, generated to evaluate the effectiveness of the classifier in the earlier steps of the evaluation stage, was passed into a custom backtesting library that would feed each days data one at a time to the trained classifier and then make trades based on the recommendations issued by it – thus, mimicking a real-world scenario of consuming data and making trades.

As the purpose of the backtesting library was merely to evaluate the effectiveness of the classifier and the process of building an effective classifier was the actual focus of this study, the library and its function were kept fairly straightforward where testing was carried out only for one equity at a time and only one share could be held or traded at any given time. Though the library enabled both buying and shorting of equities and even broker fees were accounted for where it was set to a standard cost of $0.01 per transaction. Moreover, as the classifier predicts the direction of price movements for a change of 1% in either direction (as per Table 2), the 'stop-loss' and 'take profit' were also set to 1% because when the classifier buys an equity it predicts that there shall be an increase of at least 1%, thus when the price does start increasing, if it does, then it makes sense to exit the position once it reaches that mark. This is achieved by the take profit parameter. Whereas on the other hand if the price drops by 1% when it predicted otherwise, we know the classifier made the wrong trade and to minimise any losses one should leave that position as soon as possible.

In action, the algorithm makes trades based on the rules given in Table 5 and for its very first trade, though not depicted in the table below for simplicity, it's put in default starting position where it chooses to buy or sell exclusively based on what the classifier predicts as at that point there's no previous trade that has taken place.

**Table 5.** Actions taken by the backtesting algorithm based on the previous trade and signal generated by the classifier for the current day

|  |  | Previous Trade | |
|---|---|---|---|
|  |  | Buy | Short/Sell |
| Current Signal | Buy | Hold and check stop conditions | Buy/Short Buy |
|  | Sell | Sell | Hold and check stop conditions |

Though due to the labour-intensive nature of feeding data for only one equity at a time, the backtesting was carried out only for the very popular FAANG equities (Facebook, Apple, Amazon, Netflix, Google) as a proof of concept and the results of the same have been shared in section 4.4. This way the backtesting algorithm runs over ~250 days of data for each of the equities, which is the size of the testing set generated earlier. Lastly, as every trade is made, corresponding profit and loss is calculated on it and a tally of total profit is maintained as shared in Table 7.

## 4 Results

Last 20 years of data (1999-04-04 to 2019-04-04) was analysed that consisted of 28 fundamental and technical indicators (Table 1).

### 4.1 Result 1: Model predictions are most effective for the 10$^{th}$ day in the future

Six classifiers (Table 3) were trained and tested on the entire S&P500 dataset with 28 features. Random Forest outperformed all the other classifiers with the maximum micro-weighted F1 score aka accuracy of 83.62% on 10$^{th}$ day and in second place came the K-Nearest Neighbour with an accuracy of 76.82% on day 10. Interestingly, the ability to predict the price of a stock of the two models increased as the days passed i.e. based on the "properties" of an equity on a certain day our model made more accurate predictions for 10$^{th}$ days into the future as compared to 5$^{th}$ day. Figure 3 depicts the performance of models in predicting each of the label class from Day 1 to Day 10. Each bar represents the performance of a model on a given day and bars are grouped by the model type.

For the Buy signal predictions, the Random Forest classifier predicted buying opportunities with much greater precision when compared to other classifiers i.e. if the random forest classifier predicted that an equity was a 'buy' then the chances of it truly being a 'buy' were a lot higher. It is important to be mindful here that being precise for buy didn't mean it predicted all the buy opportunities, it just meant out of the ones it was predicting 85% on day 10 were true buys and to build a reliable system that's a property we need. The classifiers like SVM and Naïve Bayes have failed to predict any buys on some of the days which highlighted their failure on the particular dataset hence the missing bars in Figure 3.

For the Sell Signal predictions, the Random Forest classier, performed poor compared to few other classifiers (Figure 3). The Random Forest recall was only 88% of the sell signals correctly when the other classifiers except K-Nearest Neighbour showed a recall of 100%. Though this should not be viewed as a shortcoming because classifier like SVM and Naïve Bayes didn't predict any buy signals on certain days meaning all the signals that were produced were sell signals and thus the recall value of 1 for sell signals was merely a consequence of not predicting any buy values.

Overall, the Random Forest classifier showed great results across the board with the best values on the day 10: 85% precision for buy signals, 88% recall for sell signals and an F1 score of 83.62%.

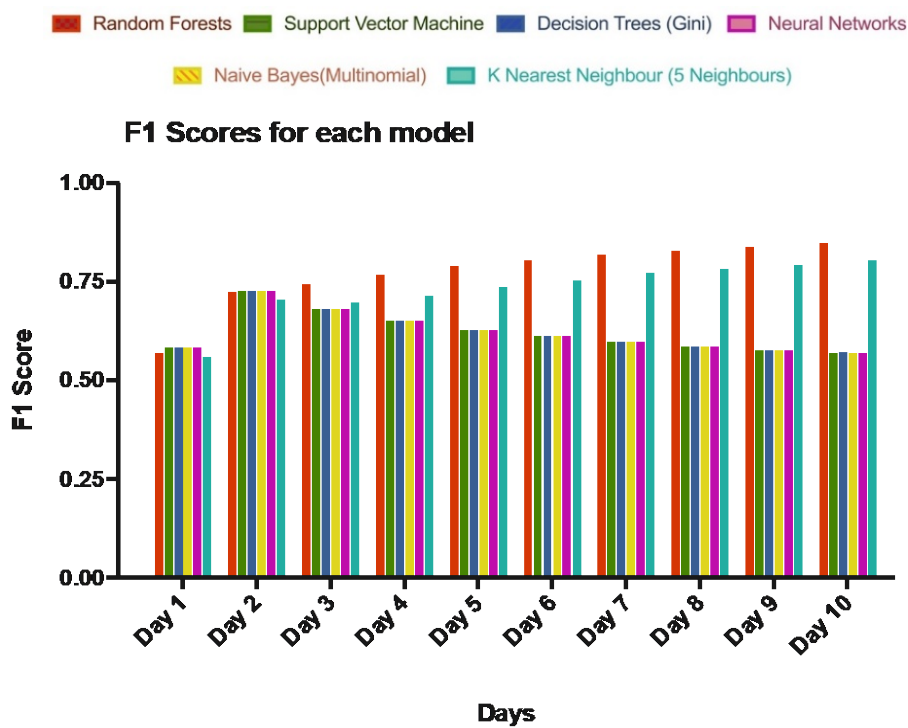

**Figure 2.** Performance of various classifiers. Most models make the most accurate predictions for 10th day ahead in the future.

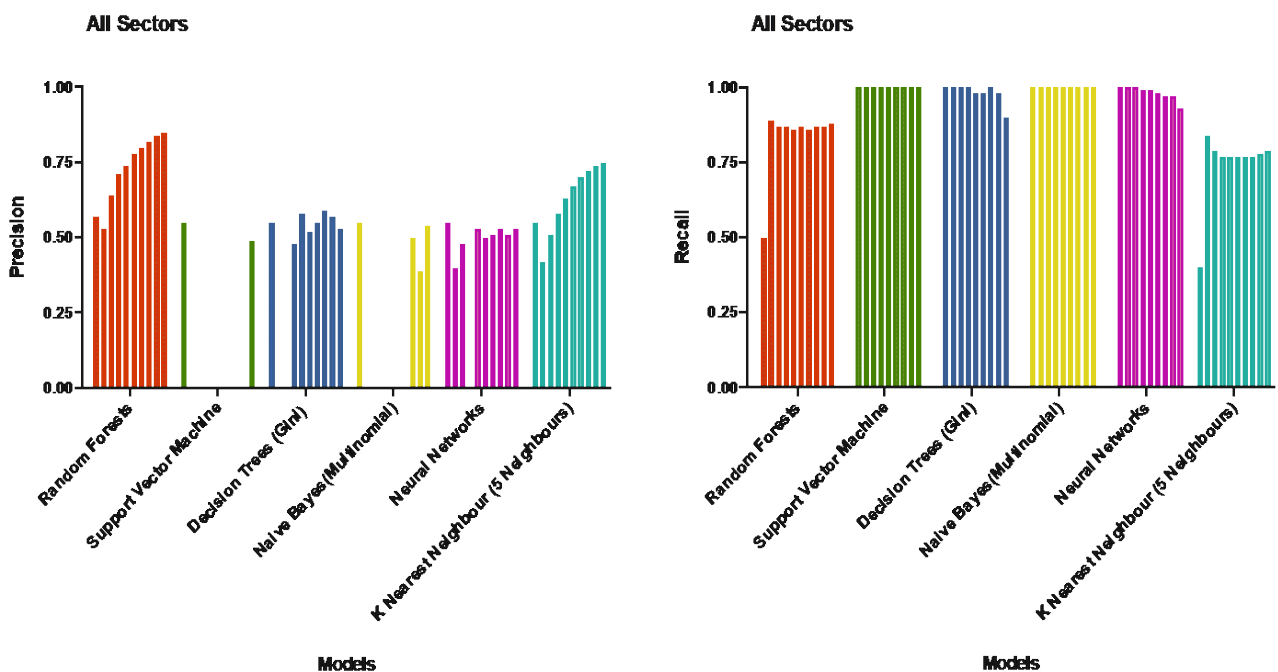

**Figure 3.** Precision for Buy signals and Recall for Sell signals of various classifiers over the S&P500 dataset. Bars are ordered from Day-1 to Day-10 from left to right.

#### 4.2  *Result 2: Sector-wise analysis didn't improve performance*

Studies by [9,10] promoting contextual analysis, one would believe that grouping equities belonging to the same industry sector would result in improved performances, however, our analysis showed this wasn't the case. Figure 4 shows that the trend remained the same as for the overall data set from Figure 3. Moreover, not just the trend, even the absolute values have remained relatively the same and with no major differences across the board. There have been instances of increases as well as decreases. For example, in the Energy sector, the accuracy of the random forest increased to 84.28% on day 10 but for Healthcare, it dropped to 82.96%.

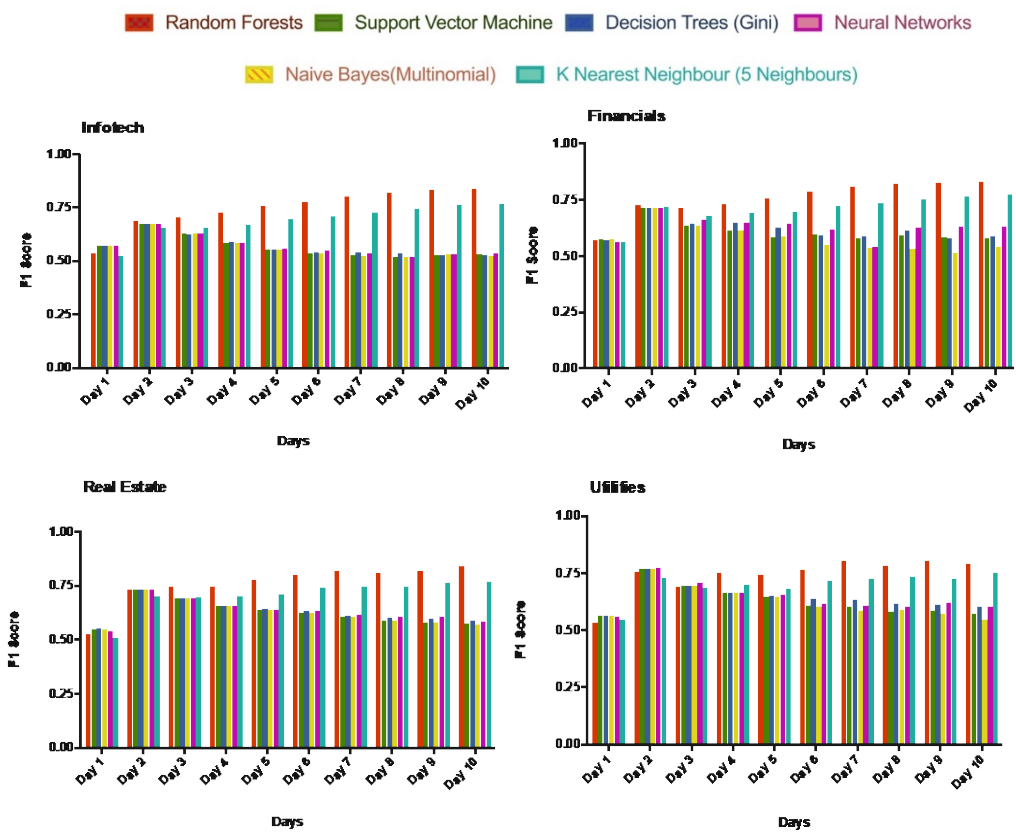

**Figure 4.** Performance of classifiers in a few selected sectors

#### 4.3  *Result 3: Just 6 indicators can be used to predict the price movement of stocks to a high degree*

We further investigated the possibility of dimensionality reduction by performing Principal Component Analysis (PCA). The first 6 components captured more than 90% of the variance while the first 15 components captured more than 99% of the variance in the data as shown in Figure 5.

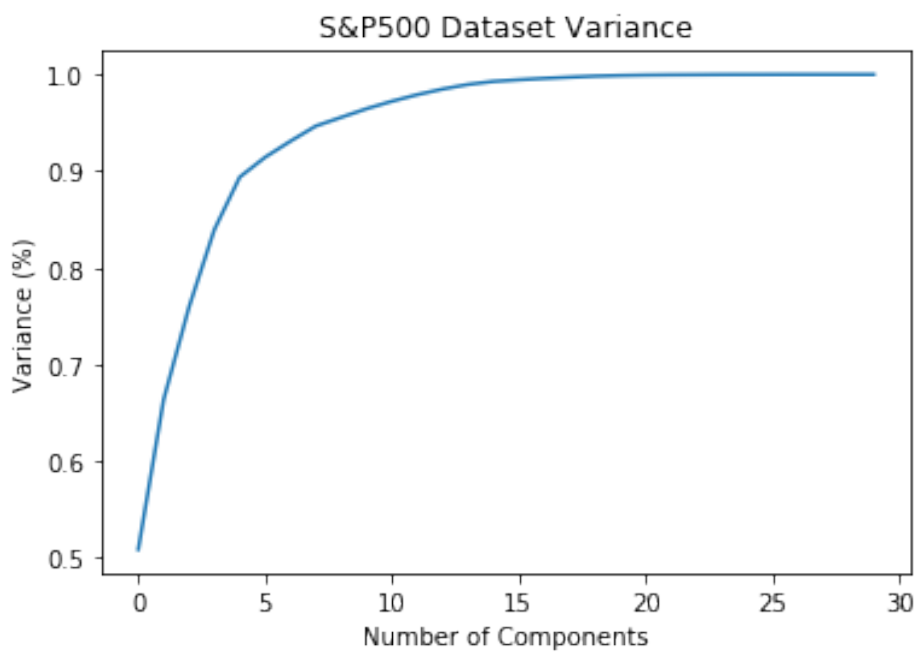

**Figure 5.** Variance captured by principal components

Since the first 6 PCA exhibited most variance in the data we further investigated which were the most important features in these components by looking at their eigenvalues against each PCA. The eigenvalue could have an absolute value from 0 to 1, so in this experiment, if the feature eigenvector had a value 0.1 or greater for a principal component then it was considered to be making a "valid contribution" to a given component. Furthermore, a feature contributing to the PC-1 was more valuable, in terms of variance representation than to a feature contributing to PC-6. Thus their occurrences were been weighted for 6 to 1. For a feature making a valid contribution to PC-1 was give 6 points, to PC-2 was given 5 and so on until a feature present in PC-1 was given a single point. The results of this process were summed up in Table 6. The column "Occurrences" was a metric, stating out of the 6 principal components (PC) how many principal components did a feature make a valid contribution to.

From Table 6, it was clear that the features that come from analyst rating, "Total Hold Recommendation" was the most significant feature that contributing in the first 5 PCA, following by "Total Buy" and "Total Sell" recommendation by the market analysts. Of 28, 6 features were not listed in Table 6 as they were considered to be not contributing to the first 6 PCA since their eigenvalue was lower than 0.1.

**Table 6.** The occurrence of each feature in the first 6 PCA components

| Feature | Occurences | Weighted Occurence |
| --- | --- | --- |
| TOT_HOLD_REC | 5 | 20 |
| TOT_BUY_REC | 4 | 17 |
| TOT_SELL_REC | 4 | 16 |
| TOT_ANALYST_REC | 3 | 11 |
| hold_percent | 3 | 11 |
| sell_percent | 3 | 11 |
| BEST_TARGET_PRICE | 3 | 10 |
| PX_OFFICIAL_CLOSE | 2 | 9 |
| EQY_REC_CONS | 2 | 9 |
| BEST_ANALYST_RATING | 2 | 9 |
| BEST_EPS | 2 | 9 |
| BEST_EPS_LO | 2 | 9 |
| BEST_EPS_HI | 2 | 9 |
| CUR_MKT_CAP | 2 | 7 |
| HISTORICAL_MARKET_CAP | 2 | 7 |
| BEST_CAPEX | 2 | 7 |
| BEST_CAPEX_LO | 2 | 7 |
| BEST_CAPEX_HI | 2 | 7 |
| buy_percent | 1 | 6 |
| SHORT_INT | 2 | 5 |
| SHORT_INT_RATIO | 1 | 1 |
| RETURN_ON_ASSET | 1 | 1 |

*4.4  Result 4:   The Randomforest classifier generates strong results in scenarios mimicking real life*

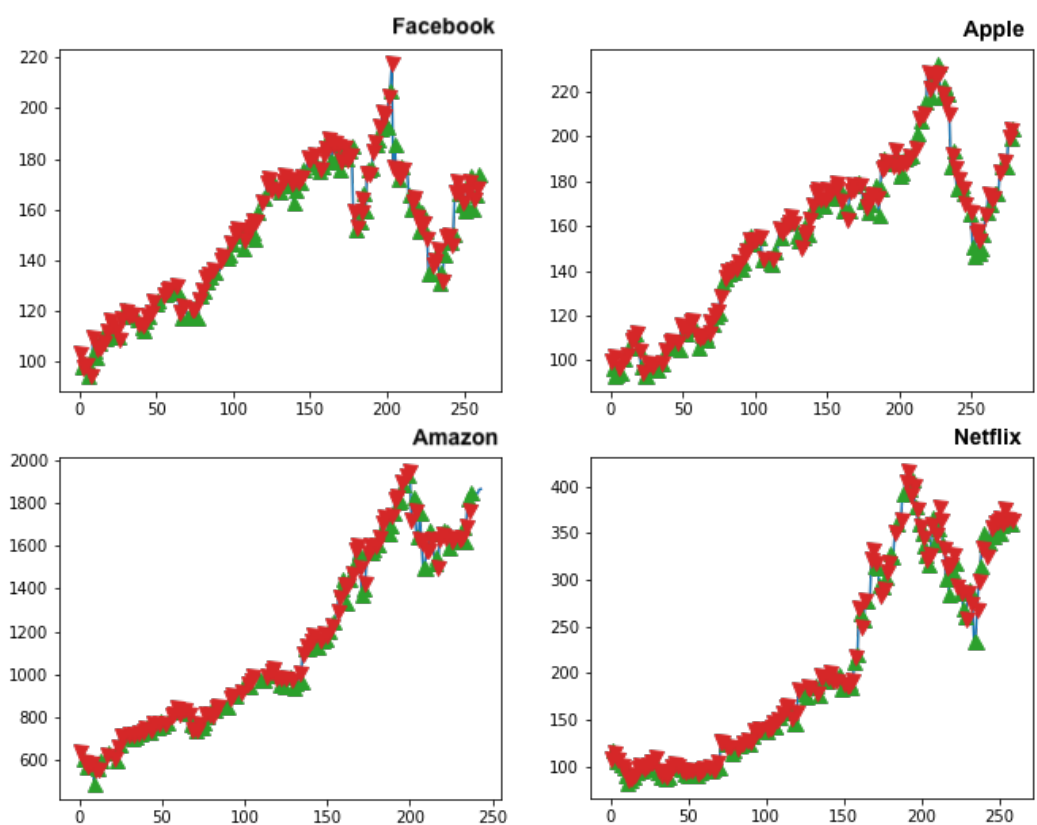

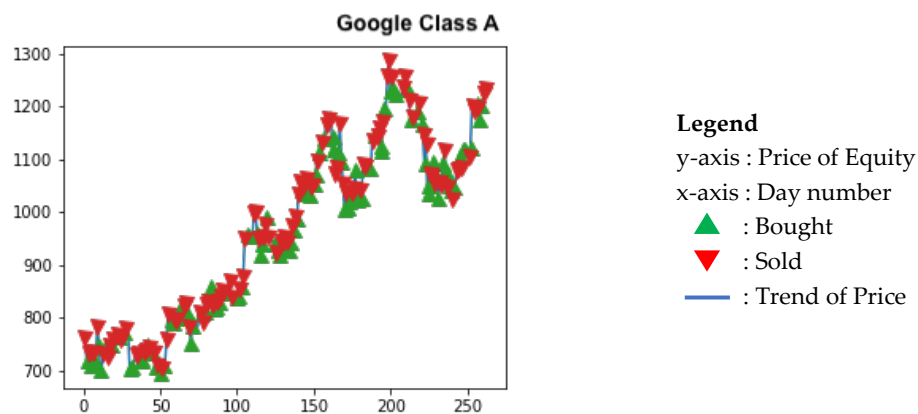

**Figure 6.** Visual representation of the transactions made during backtesting of the FAANG equities

**Backtesting:** We devised a simple backtesting system to test the trading system, the code is available at https://github.com/mkhushi/mkbacktester. We tested trading on five major S&P500 stocks known as FAANG (Facebook, Apple, Netflix, Google). The transaction cost is set to be $0.01 per share and we traded a single share at ever opptunity of buy or shorting the stock as show in Figure 6. As shown by the statistics in Table 7, the random forest classifier with the 6 indicators as identified by from the principal component analysis, was able to generate very strong returns during backtesting. Despite the FAANG being only a small share of the S&P500, it still speaks to the ability of the classifier to effectively predict price movements and generate profits. Such results were seen with a very simple trading strategy and an elementary backtesting algorithm. With improvements to the backtesting algorithm coupled with the use of other common trading strategies, this classifier can prove to be an effective foundation of a more complex system.

**Table 7:** Profits generated when backtesting on test data for the FAANG equities

| Equity | Profit generated | Price of equity at the beginning of the term | Return Percentage |
|---|---|---|---|
| Facebook | $85.63 | $102.97 | 83.16% |
| Amazon | $479.30 | $636.99 | 75.25% |
| Apple | $87.30 | $99.96 | 87.34% |
| Netflix | $118.12 | $107.66 | 109.72% |
| Google | $466.83 | $761.53 | 61.30% |

*4.5 Comparison with other studies*

Kamble [13] leveraged various classification models to predict if a set of 1000 Indian stocks would achieve the set target price of 110%, they managed to achieve an accuracy of 66.8% for the Buy Signals in their short term model and a precision of 75.82% in their long term model. On the other hand, Hargreaves et. al. [15] received an overall sensitivity of 92% and specificity of 55% implying that their system was able to successfully capture 92% of all buy signals and only 55% of the sell signals which is impressive but doesn't reflect on how many signals of the other category were incorrectly captured also, this reflects a possible bias of the system towards the buy signals.

Dai [47] modelled stock price to predict for both short and the long term i.e for the next day or the next n-days respectively. Different types of learning algorithms such as logistic regression, quadratic discriminant analysis, and SVM were utilized in price modelling on historic data ranging from Sept. 2008 to Aug. 2013. The highest accuracy achieved in the next-day prediction model was 58.2%, while the accuracy can be as high as 79.3% in the SVM based long-term model with a time window of 44. This was further improved to 96.92% with a time window of 88 days.

Considering the above studies which tackle the problem of stock prediction, one can see that an overall accuracy of 83.62%, 85% precision for buy signals, 88% recall for sell signals certainly means that the use of a combination of technical and fundamental indicators mentioned in Table 1 can improve the performance of the system. We labelled stocks based on the 1% increase/decrease in the price; with a precision of 85% on 10th day, there is potential for an investor to make 25% annual returns on investment, given that there are 250 working days in the US market.

## 5 Discussion and Conclusion

Predicting the price of an equity is a complex task. There, have been multiple studies in this area, each achieving success in their own way. This study combines the 2 schools of thought to identify a set of fundamental and technical indicators that could successfully predict the price movement of equities on the S&P500 and consequently be used to generate excess returns on the stock market.

Overall, our model showed high promise with strong numbers in all regards including accuracy, precision and recall. Specifically, our model's recall of 100% means that the model performed really well for sell signals which could be used for shorting strategy. A shorting strategy is a strategy in which trader borrow and sell in a hope that the equity of the equity we decrease in future when it could be bought to settle the trade. This strategy is considered riskier and is usually adopted by experienced traders, however, an appropriate stop loss with the confidence in machine learning along with the context of our study's results traders can be assured of a profit-making strategy. On the other hand, we have achieved a good precision of 85% on predicting the possibility of a stock's price going up by 1% on 10th day. This means that though the model missed a few buying opportunities, the ones that were predicted were true buys. In our opinion, this property is important for investors as they wouldn't mind to miss out on a few opportunities as long as they are comfortable and know that the stock they've invested in is indeed going to give them at least 1% profit. This also means that for 250 US working days there is a possibility of generating a minimum of 25% annual return (1% minimum gain on every 10th day) on investment which beats the S&P 500 annual return.

The PCA analysis showed 15 components to be representative of ~ 99% variance in the data, suggesting that though most of the 28 features were contributing towards the predictions yet it was these 15 forming the foundation of the data. By aggregating based on eigenvalue, we further identified a subset of 6 features that contribute most in the PCA-components and could predict stock prices almost as well as the original 28 features. Through this, we also established that the indicators 'Total Hold', 'Total Buy' and 'Total Sell Recommendation' which come from the analyst recommendations were highly significant in predicting the stock prices.

Finally, during the backtesting of the classifier, it generated promising returns on the initial investment. Though we acknowledge that it was carried out on a small set of equities due to the manual nature of the task, but with more sophisticated backtesting algorithms and combination of other common trading strategies, the positive returns should hold even for other kinds of equities in the S&P500, if not simply now. Another factor for future research is that our model did not consider the effect of buying of selling on the equity price movement, especially in the case of large funds whose actions of buying or selling immediately change the market dynamics. However, we still believe that our model is immediately useful for a small investor and the focus of the study was always on long-term fundamental investing and not based on the patterns of the trading chart.

Finally, with rapid improvements in the tools offered by Machine learning and Deep Learning, this is an ever-evolving field and our quest to understand how the markets work and identify trends is never-ending. Going forward, the results from this study could be tested in a different market like The Bombay Stock Exchange or the Shanghai Stock Exchange or the results could be further drilled down to identify the most optimal subset. In summary, we provide a complete workflow which is very easy to understand and implement for a real-trading system and our rigorous testing shows that the system has a potential to gain excess returns compared to the usual S&P 500 annual returns.

**Funding:**

This research received no external funding.

**Acknowledgments:** The data was analyzed at a Bloomberg Finance L.P. Terminal located at The University of Sydney.

**Conflicts of Interest:**

The authors declare no conflict of interest.